\documentclass[11pt]{article}
\usepackage{graphicx}
\usepackage{amssymb}
\usepackage{amsmath}

\newcommand{\mix}{\chi}
\newcommand{\mixe}{\chi_\mathrm{eff}}
\newcommand{\OP}{\omega_\mathrm{P}}

\newcommand{\muu}{ m_{\gamma^\prime}}

\begin{document}

\begin{flushright} 
DESY 08-061
\end{flushright}

\vspace{20pt}

\begin{center}

{\Large \bf THE LOW ENERGY FRONTIER:\\ probes with photons} \\[1.5ex]

\vspace{30pt}

{Javier Redondo}

{\small \em
Deutsches Elektronen-Synchrotron DESY,\\
Notkestrasse 85, D-22607  Hamburg, Germany}

\vspace{40pt}

{\bf Abstract}
\end{center}

\noindent
I discuss different aspects of the phenomenology of hypothetical sub eV mass particles arising in the context of extensions of the standard model. I focus on a simple extension based on an additional U(1) gauge symmetry and its corresponding gauge boson, called ``hidden photon''. Kinetic mixing with the standard photon leads to photon-hidden photon oscillations that are searched for in  laboratory experiments like ALPS at DESY. Hidden photons produced in the interior of the Sun could be also detected in axion helioscopes like CAST at CERN and could play an interesting role in late cosmology, where the presence of additional feebly interacting relativistic particles seems to be favored. All these effects disappear as the hidden photon mass decreases, allowing phenomenologically large kinetic mixings. However, in this case such a hidden photon will even play a role in gauge coupling unification.

\setcounter{page}{0}
\thispagestyle{empty}
\newpage

\emph{In the days of exploring the TeV frontier, are we leaving something behind us?}
\\
\\
It is a common opinion, and we will find numerous examples of it in this volume, that the standard model (SM) of particle physics is not completely satisfactory to describe certain aspects of nature. 
Extensions of the SM invoked to cure their diseases include generally many additional symmetries and fields. The corresponding particles have generally masses arranged to lay beyond the reach of our collider experiments (or just around the corner), namely beyond a TeV. 
It is clear that if these additional particles are very massive we have little chances of discover them in colliders, and we should rely on low energy precision experiments. 
\emph{But they could be additional light particles}. On general grounds, low masses are related to some symmetry that prevents high radiative contributions from larger mass scales. 
It is clear that the knowledge of these hypothetical low energy particles will provide us with an understanding of their related symmetries, and guide us through the difficult task of extending the standard model to describe particle physics up to arbitrarily high energies.

Of course, when these particles couple directly to the SM its existence is severely constrained from laboratory searches and our current understanding of astrophysics and cosmology. However, \emph{there are certain models in which the powerful astrophysical constraints are evaded} \cite{Masso:2006gc
}.

\section{Massive Hidden photons and the ``meV valley"}

In this contribution I focus on one of these models, whose only addition to the SM lagrangian consists in a new U(1) gauge symmetry and its corresponding gauge boson, here called ``hidden photon". The SM fields are assumed to be uncharged under this new gauge group, but nevertheless they can still interact with the hidden photon through \emph{kinetic mixing} with the standard model photon. Therefore we will consider the low energy effective lagrangian
\begin{equation}
\mathcal{L} =
-\frac{1}{4}F_{\mu \nu} F^{\mu \nu}
 - \frac{1}{4}B_{\mu \nu} B^{\mu \nu}
+ \frac{\sin\chi}{2} B_{\mu \nu} F^{\mu \nu}
+ \frac{\cos^2\mix}{2} \muu^2 B_{\mu} B^{\mu},
\label{LagKM}
\end{equation}
where $F_{\mu\nu}$ and $B_{\mu\nu}$ are the photon ($A^\nu$) and hidden photon ($B^\nu$) field strengths.
The dimensionless mixing parameter $\sin\chi$ can be generated at an arbitrarily high
energy scale and does not suffer from any kind of mass suppression from the messenger particles
communicating between the visible and the hidden sector. This makes it an extremely powerful probe
of high scale physics. 
The construction outlined here arises quite naturally in extensions of the SM based on string theory, 
where values in the range $10^{-16}\lesssim\chi\lesssim 10^{-2}$ can be expected \cite{Dienes:1996zr
}.

The most prominent implication of the kinetic mixing term is that photons are no longer massless
propagation modes. The kinetic mixing term can be removed by changing the basis $\{A,B\}\rightarrow\{A_{_R},S\}$,
where $A_{_R}=\cos\mix A$ is a renormalized photon field and $S=B-\sin\mix A$ is the state \emph{orthogonal} to it, and therefore completely \emph{sterile} with respect to electromagnetic interactions. 
The renormalization is typically unobservable and will be discussed in section 2. In this section we use $A=A_{_R}$.
In the $\{A,S\}$ basis the kinetic term is diagonal but kinetic mixing has provided an off-diagonal mass term which produces $A-S$ (vacuum) oscillations with a probability
\begin{equation}
P_{A-S}=\sin^22\mix \sin^2\frac{\muu^2 L}{4\omega} \ \ . \label{prob}
\end{equation}
where $\omega$ is the energy and $L$ is the oscillation length. It also modifies the static Coulomb potential with a Yukawa-like contribution
\begin{equation}
V(r)=-\frac{\alpha}{r}\left(\cos^2\mix + e^{-\muu r} \sin^2\mix \right) \ . 
\end{equation}

\begin{figure}[htbp]
\begin{center}
\includegraphics[width=10cm]{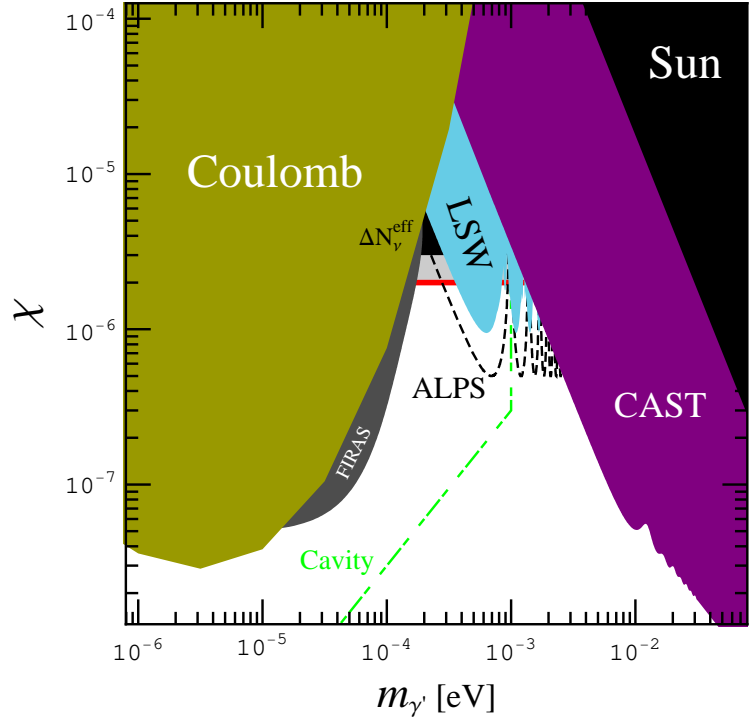}
\label{bounds}
\end{center}
\caption{
The ``meV Valley" in the mass-mixing plane of a hidden photon is bounded at low masses by searches from deviations of the Coulomb law and from seaches of solar hidden photons with the CAST helioscope at higher masses. Light-shinning-through-walls (LSW) experiments have explored the peaceful realm around $\muu\sim$ meV and an upgraded ALPS setup will penetrate even deeper in the near future. In the early universe a part of the CMB can resonantly oscillate into hidden photons contributing, as neutrinos do, to the radiation density at decoupling. Values higher than $N_\nu^\mathrm{eff}>5$ can be excluded, but a value slightly higher than 3, $N_\nu^\mathrm{eff}\simeq 3.8$ is still preferred (Red line). The precise determination of the CMB spectrum by FIRAS constraints the distortions that the creation of this hidden CMB would imprint on it. An experiment exploiting microwave cavities could be sensitive to most of the region of cosmological interest.  See the text for references.}
\end{figure}

The phenomenology of such a model has been considered by Okun \cite{Okun:1982xi} and others \cite{popov:1991
}. 
The stronger laboratory constraint comes from precision measurements of the Coulomb law \cite{Bartlett:1988yy
} and can be read off in Fig. \ref{bounds}. The sensitivity of this test  has a clear maximum at distances of the order of the centimeter, corresponding to $\muu\sim \mu$eV. For much smaller $\muu$ the hidden photon contribution is indistinguishable from that of a single massless photon, and for much higher $\muu$ the hidden photon contribution is exponentially suppressed. 

Let us now consider the astrophysical bounds, focusing on the case of the Sun. Photons of the interior of the Sun can oscillate into the sterile component that can freely escape from the solar interior removing energy that otherwise will take much longer to drain. The response of the solar structure to such an exotic energy loss is to raise the temperature of the interior so that the thermonuclear reactions can provide this extra energy. The consequence is that Hydrogen is converted much faster into Helium and the duration of the Hydrogen burning period is reduced. Studies of Raffelt and Dearborn \cite{Raffelt:1987yu} concluded that such an exotic luminosity cannot be higher than the actual visible luminosity of the Sun. Therefore, integrating eq.~\ref{prob} over the thermal distribution of photons over the solar interior will provide us with a limit \cite{popov:1991,
Redondo:2008aa} on $\mix$.

To proceed we only have to take into account an important subtlety: namely that photons in a plasma propagate as massive particles with a mass given by the plasma frequency $\OP=4\pi\alpha n_e/m_e$ with $\alpha$ the fine structure constant and $m_e,n_e$ the electron mass and number density. In such a case, the essential modification of the transition probability is the introduction of an effective mixing angle given, in the small $\mix$ approximation \cite{Redondo:2008aa}, by
\begin{equation}
\mix^2\rightarrow\mixe^2=\frac{\mix^2\muu^4}{(\OP^2-\muu^2)^2+(\omega\Gamma)^2}
\end{equation}
which strongly suppresses $A-S$ transitions, and therefore energy drain, when $\muu\ll \OP$. Here $\Gamma$ is the photon absorption rate and cuts-off the effective mixing in the resonant regime where $\muu\simeq \OP$ and the amplitude of the oscillations is maximum. The plasma frequency in the solar interior varies in the range $1$ eV $\lesssim \OP\lesssim 300$ eV (and typically $\omega\Gamma$ is smaller), so hidden photons with masses below the eV can evade the solar luminosity bound even with relatively high values of the vacuum mixing parameter $\mix$ (See Fig. \ref{bounds}).

Even a low, harmless, hidden photon flux can be detectable at earth by a suitable detector. The CAST collaboration at CERN \cite{Zioutas:2004hi
} (See the review of Silvia Borgi in this same Proceedings) operates a search for solar axions of keV energies by tracking the Sun with a $10$ m long LHC magnet, since axions emitted from the Sun can convert into photons by the inverse Primakoff effect \cite{vanBibber:1988ge}. Such an experiment will be also sensitive to hidden photons, with the benefit that high vacuum conditions are kept in the conversion region and thus the effective mixing angle is not suppressed. This nearly background free experiment can measure a photon spectral flux generated inside the magnet of $10^{-5}$ photons per second, cm$^2$ and keV. This number was used \cite{Redondo:2008aa} to set the hidden photon limit labeled CAST in Fig. \ref{bounds}. A recent paper \cite{Gninenko:2008pz} has point pointed out that considerable improvement can be achieved by measuring hidden photons of lower energies $\sim$ eV where the flux is maximal since it mostly comes from the external shells of the Sun where the electron density (and hence the plasma frequency) is smallest.

The Coulomb and CAST limits leave a valley in the allowed parameter space around the suggestive mass scale of $\muu\simeq$ meV. Since photon-hidden photon oscillations are resonant when a plasma is present such that $\OP=\muu$, it would be advantageous to find environments with a huge number of photons and electron densities $\sim 10^{15}$ cm$^{-3}$.
These conditions are found in the early universe when the temperature is of order $\sim$ keV, i.e. \emph{after big bang nucleosynthesis} (BBN) but \emph{before the cosmic microwave background} (CMB) formation. In such a scenario a fraction of the photon background will be resonantly converted into hidden photons, forming a hidden cosmic microwave background (hCMB) \cite{Jaeckel:2008fi}. 
This hCMB decouples much before than the standard CMB and from that moment on \emph{mimics the effect of additional neutrino species}, $N_\nu^\mathrm{eff}$. Since some of the CMB photons disappear, the baryon to photon ratio $\eta$ measured at decoupling also increases with respect to the value suggested by BBN. 
Therefore we can bound $\sim$ meV hidden photons from the agreement of the values of $N_\nu^\mathrm{eff}$ and $\eta$ provided from BBN and CMB physics \cite{Jaeckel:2008fi}. The CMB observations, combined with large scale structure data (LSS), slightly prefer \cite{Ichikawa:2007fa
} $N_\nu^\mathrm{eff}>3$ but both frameworks can be made to coincide within the quoted errors. The preference of a high $N_\nu^\mathrm{eff}>3$ is supported by the SDSS and Ly-$\alpha$ data and might be likely due to systematics \cite{Hamann:2007pi}. However, even in \cite{Hamann:2007pi} where a more careful treatment of the bias parameters is included, values slightly higher than 3 are still preferred, with a best global fit of $N_\nu^\mathrm{eff}=3.8^{+2.0}_{-1.6}$ ($95\%$ C.L.). It is however premature to consider that such an excess has a physical interpretation in terms of new physics, but if eventually it is confirmed it may require new weakly interacting particles that are relativistic at CMB, namely sub eV particles, and hidden photons could certainly do the job. 

Note that the conservative ``suggested" excess $\Delta N_\nu^\mathrm{eff}\simeq 0.8$ corresponds to a hidden photon with $\mix\simeq 2\times 10^{-6}$. The mass should be then $\simeq 0.2$ meV to avoid distortions of the CMB Plack spectrum and the laboratory searches to be presented next. At the view of Fig. \ref{bounds} this leads us to a clear goal in the parameter space!

Interestingly, such a scenario is going to be tested in the near future in the laboratory. 
The ALPS (Any Light Particle Search) experiment at DESY \cite{Ehret:2007cm
} is currently setting up an upgraded ``light-shinning-through-walls" experiment \cite{Sikivie:1983ip} that will explore much of the relevant parameter space in the ``meV valley". The set up consists in a powerful laser beam which propagates under high vacuum conditions to end up blocked in an opaque wall. If hidden photons or any other weakly interacting low mass particles are produced before the wall they will go through it and can be reconverted after the wall in another high vacuum cavity in which a sensitive detector is placed. Some similar experiments have been already performed \cite{Cameron:1993mr,Chou:2007zzc,Robilliard:2007bq}, the most recent motivated by the recent PVLAS episode \cite{Zavattini:2005tm
} and a recent paper has interpreted them in terms of hidden photons \cite{Ahlers:2007qf}. The current ALPS proposal includes $300$ W of laser power, conversion and reconversion lengths of $\sim 6$ meters and a small background $\sim 50$ mHz. 
The results will be presented in late fall of this same year, and immediately after several upgrades will be performed, including possibly higher laser power, a new detector and ``phase shift plates" \nolinebreak \cite{Jaeckel:2007gk} to enhance the coherence  between photons and hidden photons. 

\newpage
Already with the first upgrade the ALPS experiment will be sensitive to part of the region of major cosmological interest, and will eventually cover it completely with subsequent improvements. On the long term, additional coverage could be also provided by the mentioned new solar hidden photon searches \cite{Gninenko:2008pz} or by a photon regeneration experiment using radio waves instead of laser light \cite{Jaeckel:2007ch}.

\newpage
\section{Massless hidden photons and Unification}
All effects mentioned before are lost in the probably most natural case, a \emph{massless} hidden photon. Well, not all of them. 
We have already mentioned that to get rid of the kinetic mixing and define fields with canonical kinetic terms we need to renormalize the photon field with a factor $\cos\mix$. In the low energy lagrangian considered this is harmless, since a photon renormalization is simply reabsorbed in the definition of the electric charge as usual. Even when one considers the whole standard model gauge group and allows our new U(1) gauge boson to mix with the boson of \emph{hypercharge} (kinetic mixing with non-abelian gauge fields will not respect gauge invariance) the corresponding gauge coupling $g_1$ will absorb again this factor and leave no trace in precision electroweak observables. 

\begin{figure}[h]
\begin{center}
\includegraphics[width=8cm]{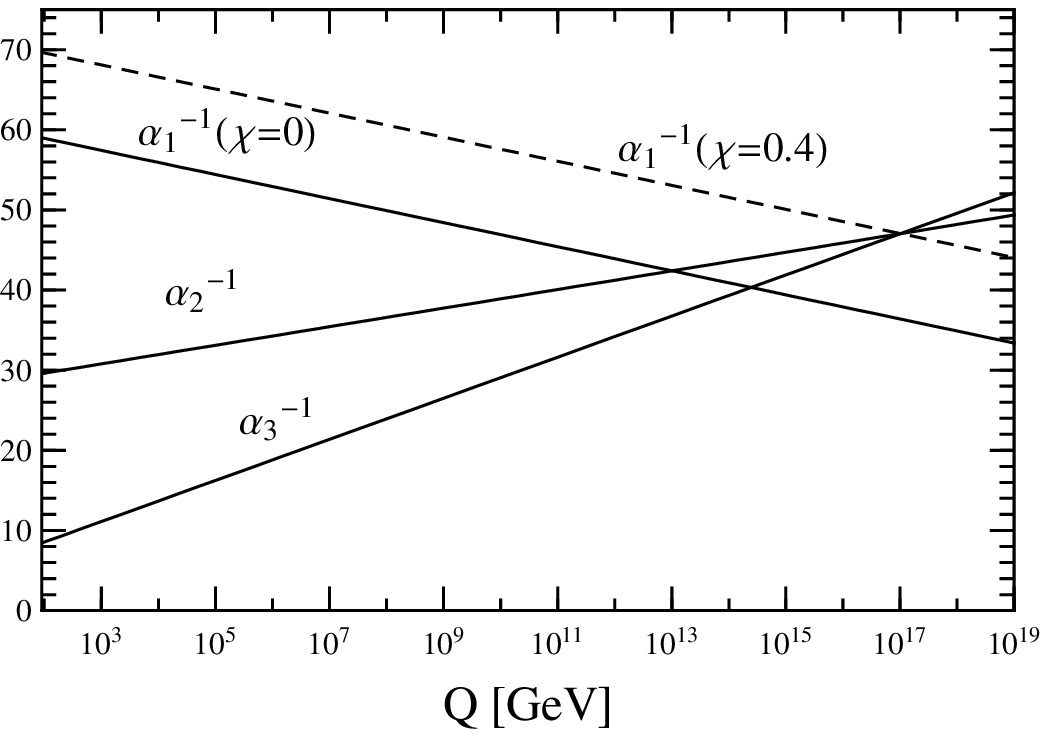}\includegraphics[width=8cm]{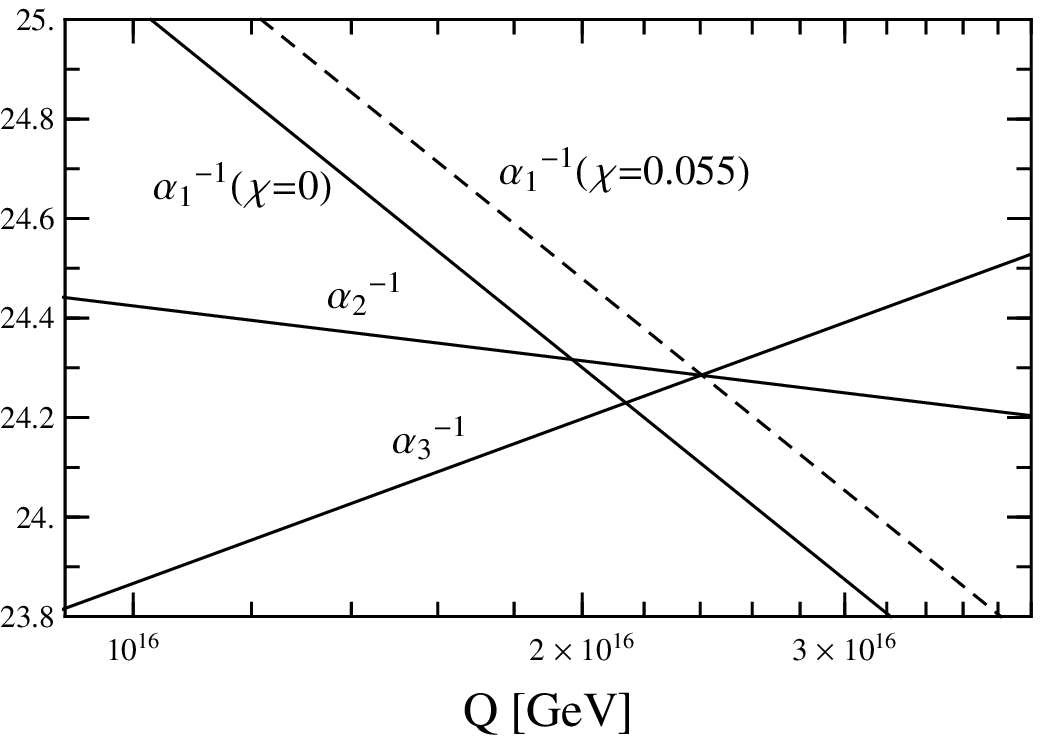}
\label{unif}
\vspace{-0.6cm}
\caption{One-loop running of the SM gauge couplings with an exotic renormalization of the hypercharge coupling $g_1$ due to kinetic mixing with an additional massless U(1) gauge boson. LEFT: standard model, RIGHT: with supersymmetry. Note that $\alpha_{1,2,3}= g_{1,2,3}^2/(4\pi)$ and $g_1$ has been normalized with the usual SU(5) factor $\sqrt{5/3}$.}
\end{center}
\end{figure}

Since this shift will only affect $g_1$ but not $g_2$ or $g_3$, it could be detectable in a theory in which there is an \emph{a priori} relation between the couplings, such as in \emph{grand unification}. In this case we shall define the unification scale by the equality of the two couplings that are unchanged $g_2(m_{_\mathrm{GUT}})=g_3(m_{_\mathrm{GUT}})$.
Note that we measure the ``renormalized" $g_1$ and this is \emph{allways larger} than the real value (the one we would expect to unify) in a factor $1/\cos\mix$, namely 
\begin{equation}
g^{\mathrm{measured}}_1=\frac{g^{\mathrm{real}}_1}{\cos\mix} \ . 
\end{equation}
Interestingly, the measured value of $g_1$ in the standard model turns out to be also \emph{larger} than the required to unify with $g_2$ and $g_3$ in a pure SU(5) model \emph{without supersymmetry}. Therefore unification could be achieved at a scale $\simeq10^{17}$ GeV (evading limits from proton decay) but being ``masked" by the exotic hypercharge renormalization due to kinetic mixing with $\mix\simeq 0.4$. We have taken values of $g_{1,2,3}$ at the Z-pole from \cite{Yao:2006px} and plotted the running in Fig. \ref{unif}.
\vspace{-0.7cm}

The case with supersymmetry (SUSY) is more complicated. 
Using the renormalization group equations at the one-loop level, a small value of $\mix\simeq 0.055$ improves the already impressive unification, but this effect is of similar magnitude than the threshold corrections of SUSY particles of $\sim$TeV masses and particles at the scale of unification. When these corrections are included the measured value of $g_1$ seems to be a bit \emph{smaller} than the required to unify perfectly \cite{Langacker:1995fk} (see \cite{deBoer:2003xm} for a recent discussion). While a more detailed study is under way \cite{Ibb}, the two possible outcomes are clear: if $g_1(m_{_\mathrm{GUT}})<g_{2,3}(m_{_\mathrm{GUT}})$ a bound on $\mix$ of order $10^{-2}$ can be set, in the opposite case, a small value of $\mix$ could be the responsible of the difference and unification could be achieved.

\providecommand{\href}[2]{#2}\begingroup\raggedright\endgroup
\end{document}